\documentclass[conference]{IEEEtran}
\IEEEoverridecommandlockouts
\usepackage{amsmath,amssymb,amsfonts}
\usepackage{algorithmic}
\usepackage[table]{xcolor}
\usepackage{graphicx}
\usepackage{textcomp}
\usepackage{xcolor}
\usepackage[nodisplayskipstretch]{setspace}
\usepackage{svg}
\usepackage[flushleft]{threeparttable}
\usepackage{array}

\DeclareMathOperator*{\argmax}{argmax}

\def\BibTeX{{\rm B\kern-.05em{\sc i\kern-.025em b}\kern-.08em
    T\kern-.1667em\lower.7ex\hbox{E}\kern-.125emX}}
\begin{document}

\title{Robust Beam Codebooks for mmWave/THz Systems: Toward a Stochastic RL Approach\thanks{This paper has been accepted for publication in the IEEE International Conference on Communications (ICC) 2026.\\ \copyright 2026 IEEE. Personal use of this material is permitted. Permission from IEEE must be obtained for all other uses, in any current or future media, including reprinting/republishing this material for advertising or promotional purposes, creating new collective works, for resale or redistribution to servers or lists, or reuse of any copyrighted component of this work in other works.}}

\author{\IEEEauthorblockN{Anouar Nechi, Rainer Buchty, Mladen Berekovic, and Saleh Mulhem}
\IEEEauthorblockA{Institute of Computer Engineering, University of Lübeck, Germany}}

\maketitle

\begin{abstract}
Millimeter-wave (mmWave) and terahertz (THz) massive MIMO systems often rely on predefined beamforming codebooks, which are usually suboptimal in Non-Line-of-Sight (NLoS) conditions and for hardware-limited transceivers. Reinforcement Learning (RL) enables adaptive, data-driven codebook design without explicit Channel State Information (CSI), but the robustness of such algorithms in practical conditions is underexplored. This paper introduces a robust multi-agent RL framework that learns beam codebooks directly from environmental feedback, eliminating the need for prior channel knowledge. Our method is well-suited for real-world deployments facing unpredictable propagation and hardware constraints. We conduct a comprehensive analysis of three off-policy algorithms, Deep Deterministic Policy Gradient (DDPG), Twin Delayed DDPG (TD3), and Soft Actor-Critic (SAC), evaluating their resilience to hardware impairments and feedback noise. Simulations show that SAC consistently outperforms deterministic methods, achieving superior beamforming gains and stability in NLoS scenarios, even under severe impairments. These results demonstrate the promise of RL-based codebook design for robust mmWave/THz massive MIMO systems.
\end{abstract}


\section{Introduction}
\label{sec:intro}
Millimeter-wave (mmWave) and terahertz (THz) MIMO systems employ large antenna arrays to mitigate path loss and ensure sufficient signal power. While analog beamforming is widely adopted for its power and cost efficiency, it restricts channel access to the RF front-end, complicating channel estimation in practical deployments~\cite{6979963}. Consequently, it often utilizes a predefined beamforming codebook (BC) for initial access and data transmission. However, such codebooks suffer from high training overhead due to the beam density required for full coverage, and their single-lobe design fails to exploit multipath diversity in NLoS scenarios. It also assumes perfectly calibrated arrays with known geometries, which results in high calibration costs and limits practicality for systems with arbitrary or imperfect hardware, especially in dynamic environments. In particular, the analog beamforming performance is strongly affected by hardware impairments and noisy feedback. Studies of 28 GHz massive MIMO systems using fixed narrow-beam codebooks have shown that while near-digital performance can be achieved in favorable Line-of-Sight (LoS) conditions, RF combining losses, phase-shifter quantization, and outdated channel-state information (CSI) significantly degrade performance in realistic environments \cite{s16101555, humphreys2019overview}. This raises concerns about the robustness of such conventional BCs. To overcome these limitations, reinforcement learning (RL) has recently emerged as a data-driven alternative~\cite{khan2023machine}, enabling the system to autonomously learn optimal beam patterns directly from environmental and hardware feedback, removing the need for explicit channel knowledge or extensive manual calibration~\cite{9610084}. It allows adaptive analog beam codebook learning directly from received power feedback without explicit CSI~\cite{9610084, 11296930}. While initial results demonstrate the feasibility of RL-based beamforming, there is no clear evaluation of RL performance under hardware impairments and noisy feedback. 

\subsection{The Role of Robustness Analysis in State-of-the-Art Beamforming}
Resilience in communication systems represents the capability to maintain acceptable service levels despite disturbances, adapt to changing conditions, and recover from failures with minimal impact on operations~\cite{ArchDesignResilientNetwork2018}. Thus, a resilient communication system can address even rare challenges, switch to a different operational mode, and offer a minimum level of service. This attribute is considered an intrinsic property of the communication system resiliency~\cite {khaloopour2024resilience} and trustworthiness~\cite{nechi2023practical}. Within this framework, robustness denotes the system’s capacity to maintain normal functioning when confronted with a specific challenge~\cite{gribble2001robustness, 10530017}. If the system fails to remain robust in a given operational mode while facing a specific challenge, it should shift to an alternate mode to preserve overall communication resilience. This shows the importance of robustness analysis of a system's operational mode facing a specific challenge.

In the domain of mmWave/THz massive MIMO systems, several challenges impact the performance of analog beamforming. For instance, a phase noise limits the performance of analog beamforming at mmWave/THz frequencies. OFDM-based analog beamforming systems using static steering beams have been shown to suffer from beamforming gain loss due to oscillator impairments, motivating reference-signal-aided phase-noise compensation techniques that preserve beam alignment under model uncertainty \cite{9195748}. At sub-THz frequencies, analysis of modular analog massive MIMO arrays operating  $\sim$140 GHz demonstrated that phase noise introduces both self-noise and inter-user interference that cannot be mitigated by increasing the number of antennas, establishing fundamental performance limits for analog beamforming at very high frequencies \cite{9419749}. Another challenge is the imperfect phase shifter that distorts beam patterns generated by idealized steering codebooks, leading to severe performance ceilings when conventional beam assumptions are used \cite{9069262}. Robust analog codebook designs have been proposed to address specific impairments, such as static beam codebooks with directional nulling for self-interference suppression in full-duplex mmWave systems \cite{9686032, 10022041}. However, these approaches remain model-based and static, requiring prior knowledge of array geometry, coupling, or impairment statistics, and they lack adaptability to changing propagation or hardware conditions. To overcome these limitations and handle these impairments, RL has emerged, and initial deployment results demonstrate significant gains in beamforming performance. However, a systematic study of the robustness of different RL algorithms, particularly under hardware impairments and noisy feedback, remains underexplored, which motivates the present work.

\subsection{Paper Contribution}
While previous works~\cite{6979963, khan2023machine, 9610084} have established the feasibility of multi-agent RL for fast codebook learning, this work systematically addresses the critical, yet largely unexplored, gap in operational robustness. Building upon our architecture in~\cite{11296930}, we introduce a novel robustness formulation and stress-testing methodology to quantify resilience against realistic physical-layer impairments. Consequently, our work expands upon~\cite{11296930} through the following key contributions:
\begin{itemize}
    \item We introduce a rigorous evaluation methodology that explicitly models hardware impairments as phase mismatches. We stress-test three off-policy algorithms (DDPG, TD3, SAC) and demonstrate that stochastic policies (SAC) are more robust to hardware impairments compared to deterministic policies (DDPG, TD3).
    \item We analyze the stability of RL-based beamforming under noisy feedback conditions. We provide a comprehensive benchmark of feedback signal corruption, establishing that the proposed robust learning strategy maintains high beamforming gain even when feedback noise reaches 40\%, identifying the upper operational limits for data-driven beamforming in practical THz deployments.
\end{itemize}
\section{RL-based Beam Codebook Learning}
\label{sec:RL}
This section highlights RL-based beamforming modeling proposed in~\cite{11296930}, and introduces the formulation of the beam codebook design as a Markov Decision Process (MDP).
\subsection{RL-based Beamforming modeling}
We consider the downlink of a mmWave/THz massive MIMO system in which a Base Station (BS) with a uniform linear array (ULA) of $M$ antennas serves single-antenna users via an analog-only beamforming architecture with a single RF chain. The BS utilizes a codebook $\mathcal{W}$ of $N$ beamforming vectors. For a given vector $\mathbf{w} \in \mathcal{W}$, the received signal $y_u$ at user $u$ is expressed as:
\begin{equation}
    y_u = \mathbf{w}^H\mathbf{h}_ux + n
    \label{eq_rx_signal}
\end{equation}
where $x$ is the transmitted symbol, $\mathbf{h}_u \in \mathbb{C}^{M \times 1}$ is the channel vector, and $n$ is additive white Gaussian noise. The beamforming vector is defined as $\mathbf{w} = \frac{1}{\sqrt{M}} [e^{j\theta_1}, \dots, e^{j\theta_M}]^T$, where each phase $\theta_m$ is selected from a discrete set of $2^r$ uniformly distributed values in the interval $(-\pi, \pi]$, with $r$ denoting the number of quantization bits. The propagation environment follows an $L$ path geometric channel model $h_u = \sum_{l=1}^{L} \alpha_l a(\phi_l)$, where $\alpha_l$ and $\phi_l$ denote the complex gain and angle of departure (AoD) of the $l$-th path, respectively. Ideally, the array response $a(\phi)$ can be modelled as:
\begin{equation}
    \left[a(\phi)\right]_m = e^{j(kd_m\cos(\phi))}
    \label{eq_arr_resp}
\end{equation}
where $k$ is the wavenumber, and $d_m$ denotes the position of the $m$-th antenna. Any perturbation renders standard CSI acquisition infeasible.

The RL system aims to maximize the SNR $\gamma_u = g_u \rho$ for user $u$, where $\rho = P_x / \sigma_n^2$ represents the transmit power-to-noise ratio and $g_u = |w^H h_u|^2$ denotes the beamforming gain. Given that the number of available beams $N$ is typically smaller than the total user population, the system first groups users with similar spatial characteristics into clusters $\mathcal{H}_s$. Consequently, the system objective is to identify the optimal codebook $\mathcal{W}^*$ that maximizes the average beamforming gain across these clusters~\cite{11296930}:
\begin{equation}
\begin{gathered}
\mathcal{W}^{*} = \argmax_{\mathcal{W}} \frac{1}{|\mathcal{H}_s|} \sum_{\mathbf{h}_u \in \mathcal{H}_s} |\mathbf{w}^H \mathbf{h}_u|^2 \\
\text{s.t}\quad w_m = \frac{1}{\sqrt{M}} e^{j\theta_m}, \quad \theta_m \in \Theta, \forall m=1,\dots,M
\label{eq_obj}
\end{gathered}    
\end{equation}
This formulation frames the codebook design as a data-driven control problem in which the agent learns optimal phase configurations directly from environmental feedback.
\begin{figure*}
    \centering
    \includegraphics[width=0.98\textwidth]{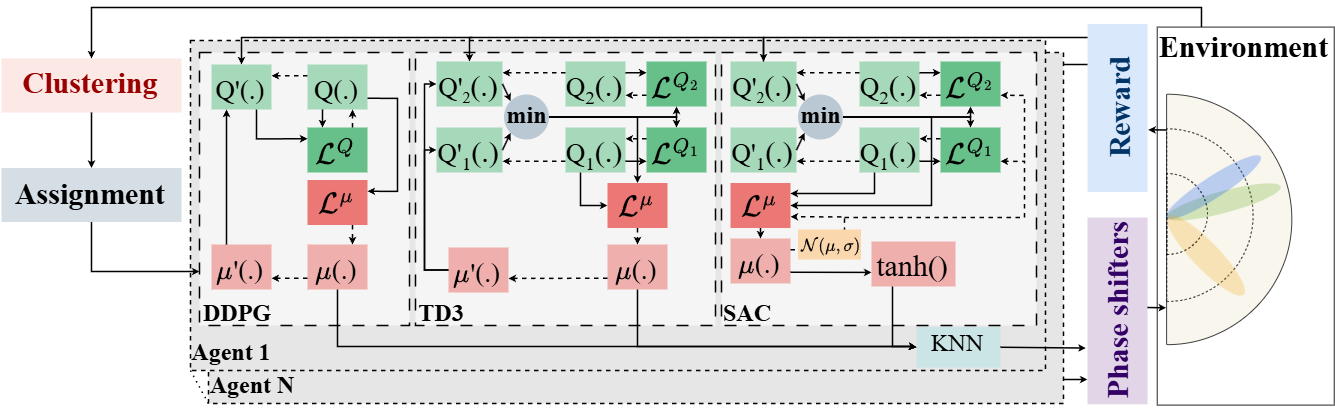}
    \caption{Proposed Multi-Agent RL Framework. The architecture decomposes codebook design into clustering and assignment phases. Individual agents optimize beam patterns using DDPG, TD3, or SAC, where continuous actions are quantized via KNN to meet hardware constraints.}
    \label{fig_RL_concept}
\end{figure*}

\subsection{MDP Formulation}
The beam codebook design can be formulated as a Markov Decision Process (MDP). Here, the RL agent interacts with the environment to optimize beam patterns solely through feedback, thereby circumventing the need for explicit channel knowledge or geometric calibration. The overall architecture of our proposed framework, including the multi-agent decomposition and the internal structure of the RL agents, is illustrated in Fig. \ref{fig_RL_concept}. The interaction is defined by the tuple $(\mathcal{S}, \mathcal{A}, \mathcal{R})$, designed to handle the constraints of analog beamformers:
\subsubsection{State \& Action Spaces} The state space $\mathcal{S}$ consists of the phase values $s_t$ of all $M$ phase shifters, allowing the agent to directly observe the hardware configuration. The action space $\mathcal{A}$ consists of phase adjustments $a_t$ output by the actor network. To ensure physical realizability in discrete phase shifters, these continuous actions are mapped to the nearest valid quantized phase using K-Nearest Neighbors (KNN).
\subsubsection{Ternary Reward} The reward $\mathcal{R}$ employs a ternary mechanism $r_t \in \{-1, 0, +1\}$ to handle noisy environment. A positive reward ($+1$) is given if the gain $g_t$ exceeds an adaptive threshold $\beta_t$ and becomes the new threshold. A neutral reward ($0$) is assigned if performance improves ($g_t > g_{t-1}$) but remains below optimal, which encourages the agent to explore promising regions without being penalized for not immediately finding the global maximum. Furthermore, relying on relative improvements ($g_t$ vs. $g_{t-1}$) rather than absolute values naturally acts as a high-pass filter, ignoring static noise biases.

\subsection{Algorithm Selection}
Three off-policy algorithms are selected to solve this MDP, categorized by their policy characteristics and resilience to hardware stochasticity.
\subsubsection{Deterministic Policies (DDPG \& TD3)} Deep Deterministic Policy Gradient (DDPG) and Twin Delayed DDPG (TD3) fundamentally rely on learning a deterministic mapping from states to optimal actions. To induce exploration during training, these algorithms add exogenous exploration noise to the policy output. In the context of analog beamforming with noisy feedback, this approach presents two critical vulnerabilities. First, the exploration noise is rigid and decays according to a fixed schedule, preventing the agent from re-initiating exploration if hardware conditions shift. Second, DDPG is prone to Q-value overestimation, where the critic network mistakenly classifies a suboptimal beam as optimal due to a transient positive spike in the noisy feedback signal.

TD3 addresses the overestimation issue by employing \emph{clipped double Q-learning}, where two parallel critics estimate the value function, and the minimum of the two is used for updates. While this provides more conservative and stable value estimates than DDPG, the underlying policy remains deterministic. Consequently, if the hardware phase mismatch causes the "true" optimal phase vector to shift slightly, a deterministic policy often fails to recover, collapsing to a local optimum that is no longer valid.
\subsubsection{Stochastic Policy (SAC)} In contrast, Soft Actor-Critic (SAC) optimizes a stochastic policy. Instead of outputting a single phase vector, the network learns a probability distribution over the action space. This is achieved by maximizing a dual objective: the expected cumulative reward and the policy entropy.

This entropy regularization is the decisive factor for robustness in hardware systems operating in the presence of impairments. In the presence of phase mismatches, the optimal beam is not a single, sharp point in the phase space, but rather a region of valid configurations. SAC’s entropy term encourages the agent to learn this entire "cloud" of high-performing actions rather than collapsing to a single point. This stochasticity acts as a natural filter for zero-mean feedback noise, allowing the agent to maintain an adaptive balance between exploration and exploitation—keeping entropy high when hardware uncertainty is significant and reducing it only when a robust, stable connection is established.
\subsection{Multi-Agent Codebook Learning}
Learning a high-dimensional codebook with a single agent is computationally intractable. We address this by decomposing the global optimization into independent sub-problems using a multi-agent architecture.
\subsubsection{Channel Clustering} 
We first group users with spatially correlated channels. Standard clustering of raw power values fails due to the "near-far" effect, in which distant users with identical angles appear dissimilar. To isolate directional features, we employ a sensing-based approach using a set $\mathcal{F}$ of $S$ random sensing beams. We transform the received sensing vector for user $k$ into a feature vector $\mathbf{u}_k$ based on pairwise power differences. For any two sensing beams $f_i, f_j \in \mathcal{F}$, the feature vector entry is computed as:
\begin{equation}
[\mathbf{u}_k]_{i,j} = \frac{|f_i^H \mathbf{h}_k|^2 - |f_j^H \mathbf{h}_k|^2}{\sum_{s=1}^{S} |f_s^H \mathbf{h}_k|^2}
\end{equation}
This normalization ensures that users are clustered solely on the basis of their angular characteristics, independent of path loss. We then apply K-means to these feature vectors to partition users into $N$ clusters.
\subsubsection{Cluster-Agent Assignment} To maximize the starting performance, we assign each newly formed cluster to the RL agent best initialized for it. We model this as a linear sum assignment problem. We construct a cost matrix $\mathcal{Z}$, where the entry $\mathcal{Z}_{n,n'} = \frac{1}{|\mathcal{H}_{n'}|} \sum_{\mathbf{h} \in \mathcal{H}_{n'}} |\mathbf{w}_n^H \mathbf{h}|^2$ represents the average gain the $n$-th agent's current policy achieves on the $n'$-th cluster. The optimal assignment is found by minimizing the total cost using the Hungarian algorithm \cite{3800020109}:
\begin{equation}
\min_{\mathbf{X}} -\sum_{n=1}^{N} \sum_{n'=1}^{N} X_{n,n'} \mathcal{Z}_{n,n'}
\end{equation}
where $\mathbf{X}$ is a permutation matrix. This ensures agents are matched with users whose policies are similar to those previously learned, minimizing retraining time.
\subsubsection{Parallel Learning} Following the assignment, the system enters the parallel learning phase. Each RL agent interacts exclusively with its assigned user cluster, updating its policy to maximize beamforming gain for that sector while avoiding inter-agent interference.

\section{New Robustness Framework for RL-based Beam Codebook Design}
\label{sec:RL_Rob}
This section introduces issues and impediments that affect the performance of the RL-based beam codebook and proposes a new robustness framework.

\subsection{Robustness Issues of RL-based Beam Codebook}
The optimization defined in \eqref{eq_obj} represents an ideal scenario. In practical mmWave/THz deployments, however, standard optimization is disrupted by two main issues.
\subsubsection{Noisy Feedback}
The feedback link can be corrupted by noise. We model such noise as additive White Gaussian Noise (AWGN) that scales with the signal strength. In this case, the agent observes a perturbed gain measurement as,  
\begin{equation}
    \tilde{g}_u=g_u + \delta= |\mathbf{w}^H \mathbf{h}_u|^2+ \delta
    \label{eq_noise}
\end{equation} 
where $\delta \sim \mathcal{N}(0, (\eta g_u)^2)$ is noise drawn from a Gaussian distribution with intensity $\eta$. Consequently, this model affects how to identify the optimal codebook $\mathcal{W}^*$ that maximizes the average beamforming gain as defined in~(\ref{eq_obj}). The proposed model captures the reality that measurement uncertainty scales with signal magnitude. 

\subsubsection{Hardware Impairments}
Second, hardware impairments that introduce unknown phase mismatches to the antenna elements. This can be modeled by modifying~\eqref{eq_arr_resp}. Unlike ideal models, we explicitly account for hardware impairments by modeling the array response $a(\phi)$ with phase mismatches:
\begin{equation}
    \left[a(\phi)\right]_m = e^{j(kd_m\cos(\phi)+\Delta \theta_m)}
    \label{eq_arr_resp2}
\end{equation}
where $ \Delta \theta_m \sim \mathcal{N}(0, \sigma_p^2)$ represents a fixed, yet unknown phase shift deviation caused by hardware impairments. This impairment renders standard CSI acquisition infeasible.

\subsection{Robustness Framework for RL-based Beam Codebook}
By modeling the noisy feedback in~\eqref{eq_noise} and the hardware impairment impacts in~\eqref{eq_arr_resp2}, we formulate the robust RL-based beam codebook problem as an empirical comparison of RL policy classes for superior robustness. The goal is to find a policy $\pi_{\theta}$ that maximizes the expected gain under such conditions (robustness issues):
\begin{equation}
    \mathcal{W}^* = \operatorname*{argmax}_{\pi_{\theta} \in \{\text{\tiny DDPG, TD3, SAC}\}} \mathbb{E}_{\Delta\theta, \delta} \left[ \frac{1}{|\mathcal{H}_s|} \sum_{h \in \mathcal{H}_s} |\mathbf{w}_{\pi_{\theta}}^H \mathbf{h}|^2 \right]
\end{equation}
This formulation compares deterministic policies (DDPG, TD3) with stochastic policies (SAC) to determine whether entropy-regularized exploration prevents the performance collapse observed when feedback is unreliable or hardware is faulty.
\section{Robustness Evaluation}
\label{sec:Res}
This section evaluates the proposed RL frameworks, focusing on their beamforming gain and robustness against hardware impairments and feedback noise.
\subsection{Experimental Setup}
We evaluate the proposed framework using the DeepMIMO dataset \cite{alkhateeb2019deepmimo} across three evaluation datasets derived from two scenarios, as detailed in Table \ref{tab_data_hyp}. The first represents an outdoor LoS scenario operating at 60 GHz, where BS index 3 is equipped with a 32-element ULA serving a grid of users. The second is an indoor NLoS scenario at 28 GHz, characterized by significant blockage, which forces reliance on multipath components. The third dataset extends the NLoS scenario by explicitly introducing hardware impairments, modeling random phase deviations that remain unknown to the agents. In all scenarios, the BS employs analog beamforming with 4-bit phase-shifters. To enable user clustering, we use 32 sensing beams to construct the feature matrix.
\begin{table}[ht]
\centering
\caption{Datasets generation \& training hyperparameters}
\begin{threeparttable}
\begin{tabular}{|l|c|c|} \hline
\multicolumn{3}{|c|}{\textbf{Scenarios}} \\ \hline \hline
\textbf{Name}             & O1\_60    & I2\_28B   \\ \hline
\textbf{Frequency}        & 60 GHz    & 28 GHz    \\ \hline
\textbf{Active BS index}  & 3         & 1         \\ \hline
\textbf{Antennas} (x,y,z) & (1,32,1)  & (32,1,1)  \\ \hline
\textbf{Users}            & 1101-1400 & 201-300   \\ \hline
\textbf{Bandwidth}        & \multicolumn{2}{c|}{0.5 GHz} \\ \hline
\textbf{Multipath number} & \multicolumn{2}{c|}{5} \\ \hline \hline
\multicolumn{3}{|c|}{\textbf{Hyperparameters}} \\ \hline \hline
\textbf{Network}          & Actor     & Critic    \\ \hline
\textbf{Replay Buffer}    & \multicolumn{2}{c|}{8192} \\ \hline
\textbf{Batch size}       & \multicolumn{2}{c|}{1024} \\ \hline
\textbf{Optimizer}        & \multicolumn{2}{c|}{Adam} \\ \hline
\textbf{Learning rate}    & \multicolumn{2}{c|}{$3.10^{-3}$} \\ \hline
\textbf{Weight decay}     & $10^{-2}$ & $10^{-3}$ \\ \hline
\textbf{$\alpha$\tnote{*} Learning rate} & \multicolumn{2}{c|}{$3.10^{-3}$} \\ \hline
\end{tabular}
  \begin{tablenotes}
  \item[*] SAC Temperature parameter.
  \end{tablenotes}
\end{threeparttable}
\label{tab_data_hyp}
\end{table}

To ensure numerical stability during training, we preprocess the channel data to mitigate the impact of inherently small signal magnitudes \cite{7160780}. We apply a global normalization using a scaling factor $\Delta = \max |[h_u]_m|$, computed over all antenna elements $m$ and all user channels $h_u$ in the dataset. This scaling is crucial to the stability of the reward mechanism, which relies on the relative magnitudes of beamforming gains.

The RL agents are implemented using a standardized actor-critic architecture in PyTorch \cite{NEURIPS2019_bdbca288}. The actor networks, taking the phase vector of dimension $M$ as input, consist of two fully connected hidden layers with $16M$ neurons each and ReLU activations, followed by a tanh output layer scaled by $\pi$ to match the phase range. The critic networks process concatenated state-action pairs, resulting in an input dimension of $2M$, through two hidden layers of $32M$ neurons.
\subsection{Codebook Learning Performance}
We first assess the baseline performance of the proposed RL algorithms in ideal hardware conditions across varying codebook sizes. Table \ref{tab_ideal_perf} summarizes the average beamforming gain achieved by DDPG, TD3, and SAC as a percentage of the Equal Gain Combining (EGC) upper bound.
\begin{table}[ht]
\centering
\caption{RL Agents Performance as (\%) of the EGC Upper Bound}
\begin{tabular}{|c|c|c|c|c|} \hline
\multicolumn{5}{|c|}{\textbf{LoS Scenario}} \\ \hline \hline
\textbf{Agent} & \textbf{4 Beams} & \textbf{8 Beams} & \textbf{12 Beams} & \textbf{16 Beams} \\ \hline
\textbf{DDPG}  & 66.1             & 88.0             & 91.7              & 95.4              \\ \hline
\textbf{TD3}   & 66.4             & 87.2             & 91.9              & 96.8              \\ \hline
\textbf{SAC}   & 67.6             & 91.4             & 94.7              & 98.0              \\ \hline \hline
\multicolumn{5}{|c|}{\textbf{NLoS Scenario}} \\ \hline \hline
\textbf{DDPG}  & 60.9             & 72.6             & 75.6              & 78.5              \\ \hline
\textbf{TD3}   & 61.5             & 72.8             & 75.4              & 78.0              \\ \hline
\textbf{SAC}   & 62.4             & 74.2             & 77.1              & 80.0              \\ \hline
\end{tabular}
\label{tab_ideal_perf}
\end{table}

In the LoS scenario, the average gain increases monotonically with the number of beams. SAC consistently outperforms the deterministic policies. For a codebook size of $N=8$, SAC achieves a beamforming gain corresponding to 91.4\% of the EGC bound, whereas DDPG and TD3 achieve lower ratios of 88.0\% and 87.2\%, respectively. This performance gap remains evident as the codebook size increases to $N=16$, at which point SAC effectively exploits the spatial resolution to reach 98.0\% of the theoretical limit, compared to 96.8\% for TD3 and 95.4\% for DDPG. 

In the NLoS scenario, the challenge is significantly greater due to multipath propagation and signal blockage. Despite this, SAC maintains its superiority. With $N=4$ beams, SAC reaches 62.4\% of the EGC limit. As the codebook expands to $N=16$, SAC continues to lead, achieving 80.0\% of the bound. This demonstrates that its stochastic exploration strategy is better suited for capturing the diverse scattering components of the NLoS channel than the noise-based exploration of TD3 and DDPG.

\subsection{Robustness to Hardware Impairments}
\begin{figure}[ht]
    \centering
    \includegraphics[width=0.95\linewidth]{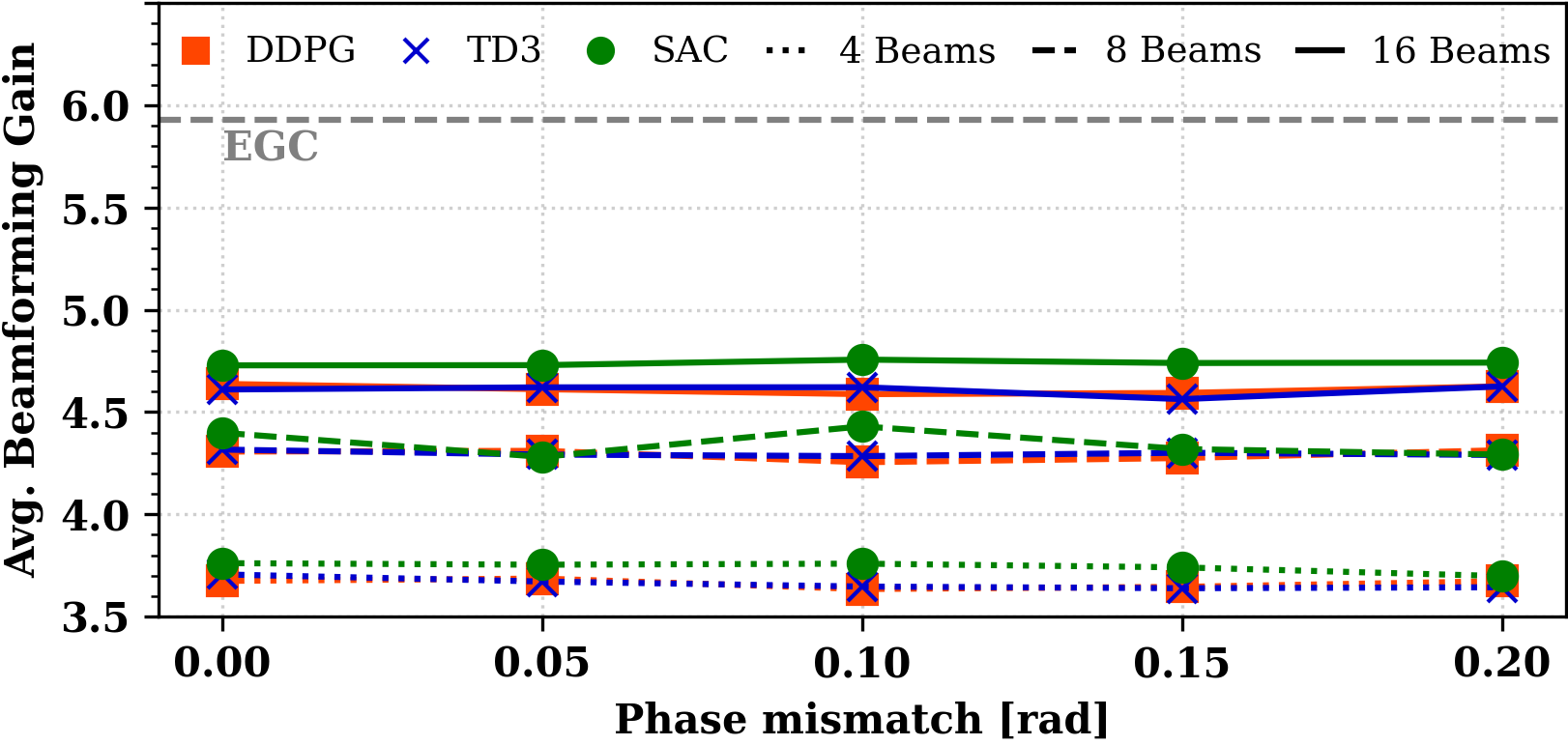}
    \caption{Average beamforming gain versus phase mismatch standard deviation $\sigma_p$ in the NLoS scenario across various codebook sizes.}
    \label{fig_Perf_HI}
\end{figure}

We evaluate the robustness of the proposed framework to hardware impairments by varying the phase-mismatch standard deviation, $\sigma_p$, from $0.00$ to $0.20$ rad in the NLoS scenario. Fig. \ref{fig_Perf_HI} presents the average beamforming gain for varying codebook sizes ($N=4, 8, 16$) under these conditions. The results demonstrate that the proposed framework is highly robust to hardware imperfections. The performance remains stable across all algorithms, even as the severity of the phase mismatch increases. Notably, SAC consistently achieves the highest gain across all configurations. This performance advantage stems from its entropy-regularized policy. The stochastic nature of SAC enables it to navigate the dynamic multi-user environment and adapt to hardware irregularities more effectively than the deterministic policies of DDPG and TD3.

An interesting behavior is observed in the 8-beam configuration, which shows slightly greater volatility than the other setups. This instability arises because the beam resolution is critically matched to the angular spread of the user clusters. In this "critical matching" regime, users frequently reside on the angular switching boundary between two adjacent beams. Consequently, minor phase mismatches induced by the hardware can trigger beam switching, leading to fluctuations in the observed gain. In contrast, the 4-beam setup provides a resolution too coarse to detect these minor shifts, resulting in stable but lower gain. Conversely, the 16-beam setup offers sufficient density to cover transitions smoothly, thereby maintaining both high gain and stability.

\subsection{Robustness to Feedback Noise}
\begin{figure}[ht]
    \centering
    \includegraphics[width=0.8\linewidth]{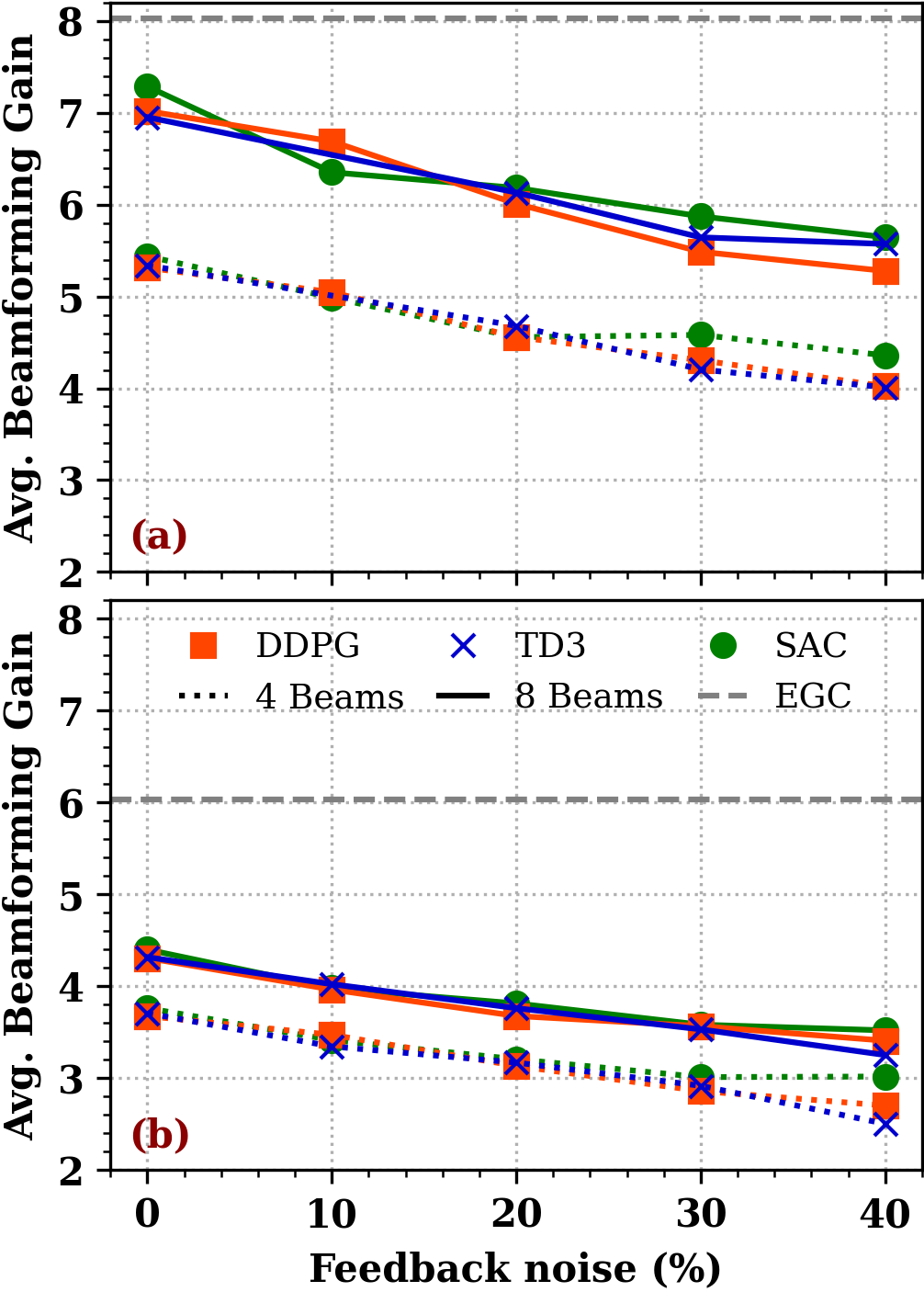}
    \caption{Impact of feedback noise on beamforming gain in LoS (a) and NLoS (b) scenarios for 4-beams and 8-beams codebooks.}
    \label{fig_Perf_FBN}
\end{figure}

In the absence of explicit channel estimation, RL agents rely solely on user feedback to update their policies. We evaluate the system's resilience by introducing Gaussian noise into the feedback link and simulating scenarios in which the reported signal strength is corrupted by measurement errors or corrupted feedback. Fig. \ref{fig_Perf_FBN} illustrates the impact of feedback noise intensity, ranging from 0\% to 40\%, on the average beamforming gain.

As anticipated, the performance of all algorithms degrades as the noise level increases, since the received gain value becomes a less reliable indicator of the true beam quality. However, the degradation rate varies significantly across algorithms. In the LoS scenario (Fig. \ref{fig_Perf_FBN}-a), SAC consistently outperforms DDPG and TD3 across all noise levels. For the 8-beam codebook, SAC maintains a gain of approximately 5.8 at 40\% noise, whereas DDPG and TD3 drop to approximately 5.2. This gap highlights the fragility of deterministic policies when the value function is corrupted; DDPG tends to overestimate the value of suboptimal actions that coincide with positive noise spikes.

In the more challenging NLoS scenario (Fig. \ref{fig_Perf_FBN}-b), the superior robustness of SAC is even more pronounced. The 8-beam configuration for SAC exhibits a much flatter degradation curve compared to DDPG. At 40\% feedback noise, SAC retains roughly 78\% of its noise-free performance. In contrast, DDPG suffers a sharper decline, losing nearly 36\% of its initial gain. This resilience is attributed to SAC's stochastic policy optimization. By optimizing a probability distribution rather than a single deterministic point, SAC effectively averages out the zero-mean feedback noise over multiple interactions. This filtering effect, combined with the proposed ternary reward structure, prevents the agent from diverging even when a significant portion of the feedback signals is unreliable.

\subsection{The Role of Adaptive Exploration in Resilience}
\begin{figure}[ht]
    \centering
    \includegraphics[width=0.82\linewidth]{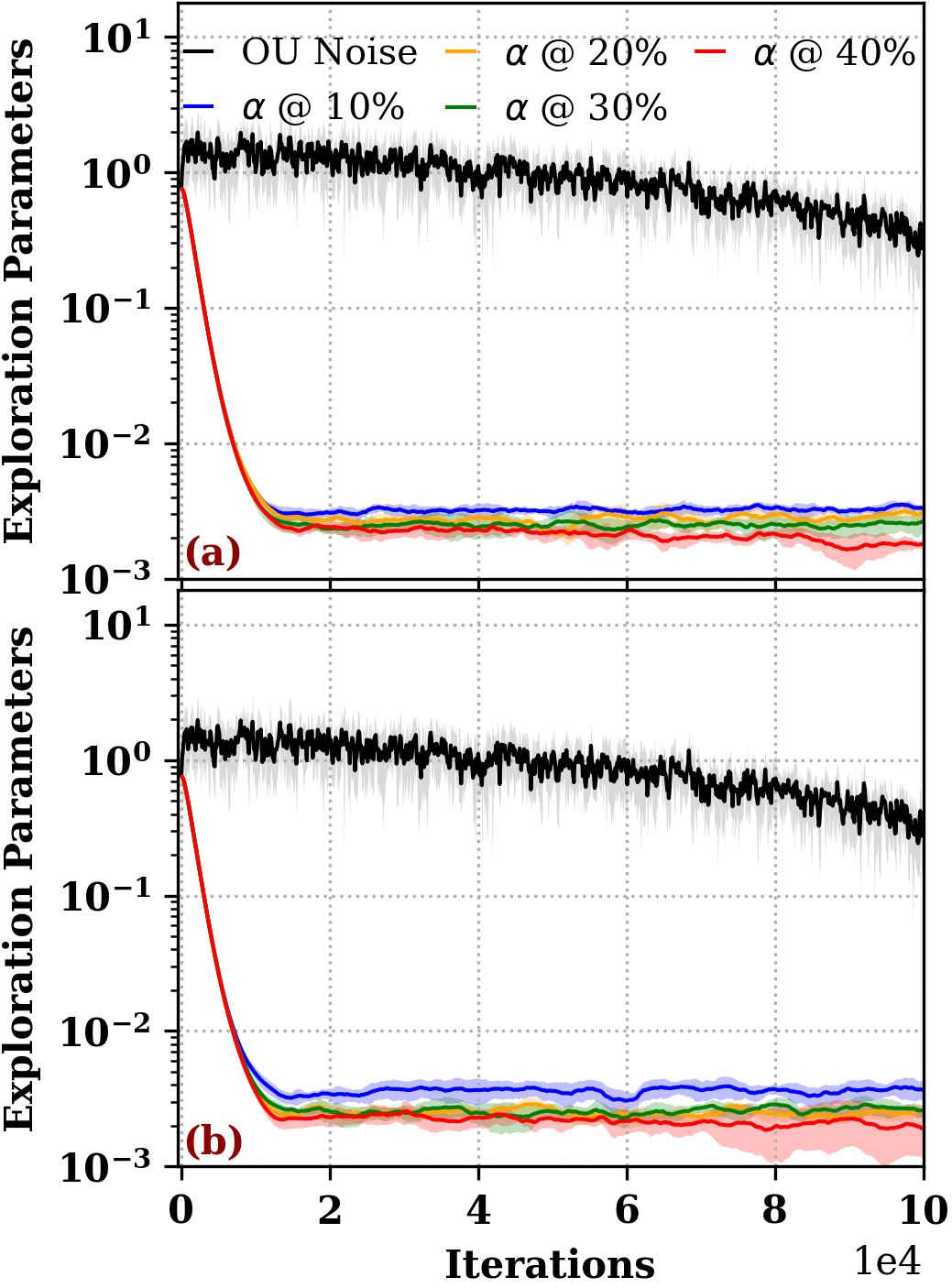}
    \caption{Evolution of exploration parameters in (a) LoS and (b) NLoS scenarios. The deterministic policies (DDPG/TD3) rely on fixed Ornstein-Uhlenbeck (OU) noise, whereas SAC adapts its entropy temperature ($\alpha$) to the environment.}
    \label{fig_exploration}
\end{figure}

To explain the performance disparity, we analyze the evolution of exploration parameters. Fig. \ref{fig_exploration} contrasts the fixed Ornstein-Uhlenbeck (OU) noise in DDPG/TD3 against the adaptive entropy temperature ($\alpha$) in SAC across varying feedback noise levels.

A critical limitation of deterministic algorithms is their rigid OU noise decay, which can maintain a high magnitude for a significant period. In noisy environments, this persistent variance prevents the agent from distinguishing exploration noise from environmental stochasticity, hindering convergence. In contrast, SAC demonstrates rapid adaptation. The entropy temperature $\alpha$ autonomously drops by three orders of magnitude (to $\approx 10^{-3}$) within 2,000 iterations. This indicates that SAC effectively transitions to exploitation by minimizing internal variance. Consequently, adaptive exploration acts as the primary resilience mechanism, filtering out high-frequency noise that destabilizes deterministic baselines.

\section{Conclusion}
\label{sec:Conclusion}
This paper addressed the robustness of RL-based beam codebook design against hardware impairments and feedback noise in mmWave/THz systems. By evaluating a multi-agent framework on DeepMIMO datasets, we demonstrated that the stochastic policy of Soft Actor-Critic (SAC) provides superior resilience compared to deterministic baselines (DDPG/TD3). Specifically, SAC’s entropy-regularized exploration effectively filters feedback noise and mitigates phase mismatches, allowing the system to maintain high beamforming gain even under severe feedback corruption. These findings establish stochastic RL as a critical enabler for resilient, hardware-agnostic mmWave/THz deployments.
\section*{Acknowledgment}
This research was partially supported by the German Research Foundation (DFG) grant 403579441, project "Meteracom".

\bibliographystyle{IEEEtran}
\bibliography{ref}
\end{document}